# The Late-Pleistocene extinction of megafauna compared with the growth of human population

Ron W. Nielsen[1]

**Abstract**. Time-dependent distribution of the global extinction of megafauna is compared with the growth of human population. There is no correlation between the two processes. Furthermore, the size of human population and its growth rate were far too small to have any significant impact on the environment and on the life of megafauna.

## Introduction

In my precious publication (Nielsen, 2017a), I have discussed results published by Barnosky (2008) who claimed that there was a strong correlation between the intensified extinction of megafauna and the growth of human population. It is both surprising and disturbing that his discussion was ever published.

I have shown that what Barnosky claimed to have been the human population *was not human population* but a set of totally meaningless numbers created by Hern (1999), the numbers so obviously fabricated that their artificial construction was easy to see. I have also shown that even these fabricated "data" did not support the postulate of human-assisted extinction of megafauna because there was obviously *no correlation between these "data" and the extinction of megafauna*. A change in the trajectory describing the extinction of megafauna was not matched by a change in the trajectory, which according to Barnosky was describing the growth of population but in fact was describing the growth of a phantom genus I called *Phasmapithecus*, the ghost population that never existed but was created by Hern (1999) and taken by Barnosky as representing the growth of human population, even thou it was perfectly obvious that it was a ghost population.

It is disturbing that science is so disrespected by people who are supposed to do scientific research. It is disturbing that science is also so disrespected in the peer-reviewed scientific journals. It is disturbing that the obviously fabricated "data" were published in a peer-reviewed journal. It is disturbing that the obviously unsubstantiated claims of Barnosky were also published in a peer-reviewed journal. It is disturbing that all this misinformation is still treated as a scientific evidence.

However, Barnosky (2008) presented an interesting time-dependent distribution of the global extinction of megafauna showing a rapid decline in the number of species between 15,500 BP and 11,500 BP (before present). I am not sure if this distribution is correct but assuming that it is correct, we can now check whether there is a correlation between the growth of real human population and the extinction of megafauna. I will compare this distribution, just as published by Barnosky (2008), with the growth of human population as described by reputable data (Biraben, 1980; Birdsell, 1972; Clark,1968; Cook,1960; Deevey, 1960; Durand, 1974; Gallant, 1990; Hassan, 2002; Haub, 1995; Livi-Bacci, 1997; McEvedy & Jones, 1978; Taeuber & Taeuber, 1949; Thomlinson, 1975; Trager, 1994; United Nations, 1973, 1999, 2013; US Census Bureau, 2017).

The discussion presented here is based on my extensive analysis (Nielsen, 2017b) of these data. It replaces the previous discussion published under the same title in 2013. I am going to show that the distribution describing the growth of human population is not correlated with the distribution describing the extinction of megafauna and, consequently, that the postulate of the human-assisted extinction of megafauna is not supported by data describing the growth of human population.

[1]AKA Jan Nurzynski, Environmental Futures Research Institute, Griffith University, Qld, 4222, Australia, ronwnielsen@gmail.com

## Contradicting evidence in population data

Figure 1 shows the growth of human population in the past 2,000,000 years compared with the best fit to the data. Details of this analysis are described in my publication (Nielsen, 2017b). Data points represent the average values of the estimates of the size of human population (Biraben, 1980; Birdsell, 1972; Clark,1968; Cook,1960; Deevey, 1960; Durand, 1974; Gallant, 1990; Hassan, 2002; Haub, 1995; Livi-Bacci, 1997; McEvedy & Jones, 1978; Taeuber & Taeuber, 1949; Thomlinson, 1975; Trager, 1994; United Nations, 1973, 1999, 2013; US Census Bureau, 2017).

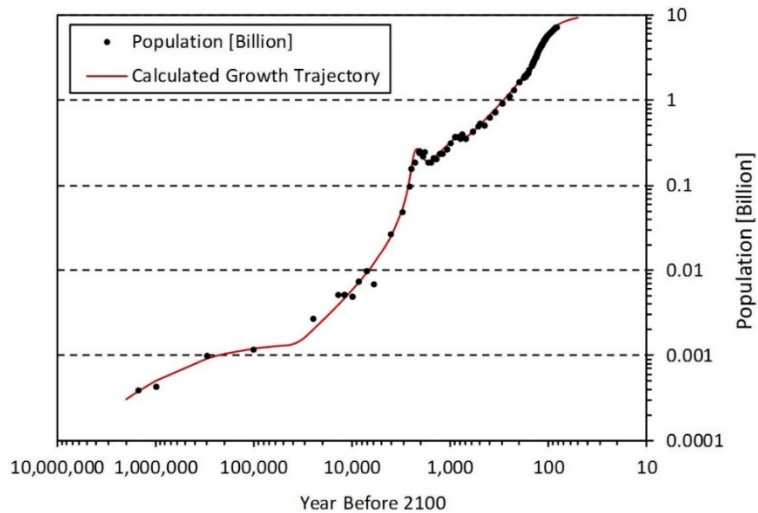

**Figure 1.** *Growth of human population in the past 2,000,000 years.*

My analysis (Nielsen, 2017b) confirms the earlier observation of von Foerster, Mora and Amiot (1960) that the growth of human population was hyperbolic during the AD time but extends it to the BC time. My analysis confirms also the observation of Deevey (1960) that the growth of human population during this long time was in three major stages but demonstrates that these three major stages are described by hyperbolic distributions. They are shown explicitly in Figure 2.

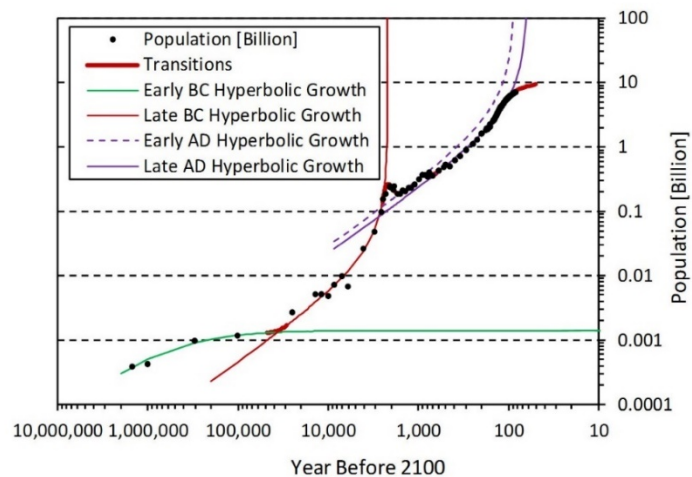

**Figure 2.** *Three major stages of growth of the world population in the past 2,000,000 years: (1) between 2,000,000 BC and 27,000 BC, (2) between 27,000 BC and AD 510 and (3) between AD 510 and present. The last stage experienced a minor distortion between around AD 1195 and 1470. This distortion caused a small shift in the hyperbolic growth.*

The three major stages of the growth of human population are: (1) hyperbolic growth between 2,000,000 BC and 46,000 BC followed by a transition (between 46,000 BC and 27,000 BC) to a new hyperbolic trajectory; (2) hyperbolic growth between 27,000 BC and 425 BC followed by a transition (between 425 BC and AD 510) to a new hyperbolic trajectory; and (3) hyperbolic growth between AD 510 and 1950, followed by a transition to a yet unknown trajectory. (For the remote time, such as 2,000,000 BC it matters little whether we express the listed years in BC, BP or as the time before 2100. The time before 2100 is needed to have positive values for the logarithmic scale of time. For the closer years, it is perhaps more convenient to express time in BC and AD scales.)

During the AD time, hyperbolic growth was slightly disturbed between AD 1195 and 1470. This disturbance coincides with the unusual convergence of five intensive demographic catastrophes (Nielsen, 2016a, 2017b). This strong and unusual event shifted slightly the trajectory describing hyperbolic growth but it also shifted it to a slightly *faster* trajectory, demonstrating the regenerating features of the Malthusian positive checks (Malthus, 1798; Nielsen, 2016b). If this disturbance was *caused* by the action of these five demographic catastrophes, then this is the only example that demographic catastrophes influenced the growth of human population. With the exception of this rare example, demographic catastrophes did not shape the growth of human population (Nielsen, 2017c).

What is essential now to notice is that the intensified extinction of megafauna between 15,500 BP and 11,500 BP (Barnosky, 2008), or between 13,550 BC and 9,550 BC (if we count the time before present from 1950), is right in the middle of the long, second stage of the hyperbolic growth of human population. The growth of population was increasing monotonically, so obviously we cannot expect any correlation between the growth of population and the non-monotonic distribution describing the massive extinction of megafauna. These two distributions are shown in Figure 3.

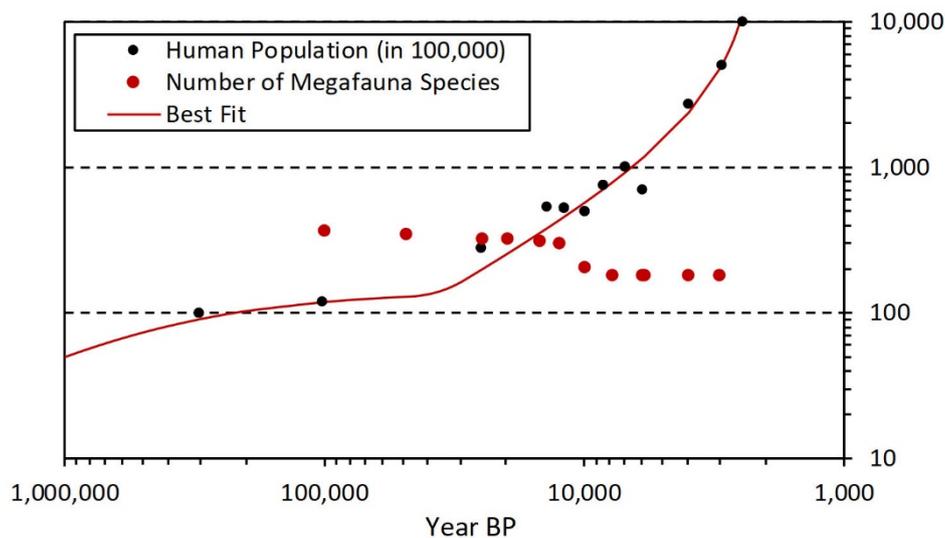

**Figure 3.** *The decline in the number of species of megafauna (Barnosky, 2008) is compared with the growth of human population. There is no correlation between these two distributions.*

Figure 3 demonstrates clearly that *there is no correlation between the distribution describing the extinction of megafauna and the growth of population*. At the time of the massive extinction of megafauna, the growth of population remained undisturbed. Extinction of megafauna did not boost the growth of human population. It is also interesting to notice that the growth of population was boosted between 46,000 BC and 27,000 BC (between 46,1950 BP and 27,1950 BP) but this boosting is not reflected in a change of the extinction trajectory. Human life had no impact on the life of megafauna and the life of megafauna had no impact on the human life. Data do not support the postulate of human-assisted extinction of megafauna.

In order to understand the possible interaction between humans and megafauna we also have to consider the *size* of human population and their *growth rate*. Was the number of people living at that time so great that they could have had a strong impact on the life of megafauna? Were they capable of not only killing a great number of these large animals but also of killing so many of them as to cause the extinction of not just one or two but approximately 160 species? The size of the *global* human population and their growth rate around the time of the extinction is shown in Table 1.

**Table 1.** *Parameters describing the growth of human population in the vicinity of the claimed (Barnosky, 2008) rapid decline in the number of species of megafauna.*

| Year [BP] | Population Size [Million] | Growth Rate [%] | Global Natural Increase [Persons/Year] |
|---|---|---|---|
| 40,000 | 1.3 | 0.0009 | 12 |
| 30,000 | 1.6 | 0.0033 | 53 |
| 20,000 | 2.5 | 0.0056 | 140 |
| 15,500 | 3.4 | 0.0074 | 252 |
| 15,000 | 3.5 | 0.0077 | 270 |
| 14,000 | 3.8 | 0.0084 | 319 |
| 13,500 | 4.0 | 0.0087 | 350 |
| 13,000 | 4.1 | 0.0091 | 373 |
| 12,000 | 4.5 | 0.0100 | 450 |
| 11,500 | 4.8 | 0.0106 | 509 |
| 11,000 | 5.1 | 0.0112 | 571 |
| 10,000 | 5.7 | 0.0126 | 718 |
| 5,000 | 15.4 | 0.0339 | 5,221 |
| 3,000 | 47.8 | 0.1056 | 50,433 |

At the peak of the extinction of megafauna, in 13,500 BP, the total global population was only 4 million. They were scattered over various parts of the world and they were supposed to have caused such a massive extinction of so many species of megafauna. The natural *global* increase of human population was then only about 350 persons per year. These few hundred individuals added to the total global population each year were supposed to balance the biomass of the killed megafauna (Barnosky, 2008; see also Nielsen, 2017a).

With such a small size of global population, people were living most likely in small groups in various parts of the world. There must have been an abundance of food for them to eat. Why should they be interested in going into all this trouble to kill megafauna, to kill these large animals, which were difficult for them to kill. However, killing them for food is one issue but killing them is such massive numbers as to cause their extinction is entirely different matter. Why would they do it? How were they supposed to do it? Who was supposed to eat all these huge quantities of meat?

If they were killing them on purpose, then there is not much sense in this sudden desire to hunt and kill these large animals. For some inexplicit reason, this desire was global. If they were killing them by accident, such as burning forests, why should they have such a universal desire to burn forests wherever they went?

## Summary and conclusions

Analysis (Nielsen, 2017b) of population data (Biraben, 1980; Birdsell, 1972; Clark,1968; Cook,1960; Deevey, 1960; Durand, 1974; Gallant, 1990; Hassan, 2002; Haub, 1995; Livi-Bacci, 1997; McEvedy & Jones, 1978; Taeuber & Taeuber, 1949; Thomlinson, 1975; Trager, 1994; United Nations, 1973, 1999, 2013; US Census Bureau, 2017) shows that at the time of the massive extinction of megafauna (Barnosky, 2008) the growth of human population was increasing monotonically by following hyperbolic trajectory. When the two distributions are compared, (the distribution describing the extinction of megafauna and the distribution describing the growth of population – see Figure 3) the presented study shows that *there is no correlation between the growth of human population and the extinction of megafauna.*

There was no rapid increase in the size of human population to correlate it with the rapid decline in the number of species of megafauna. This study also shows that throughout the entire time of the global extinction of megafauna (between 15,500 BP and 11,500 BP) and even for a long time before and after this extinction pulse, the size of human population was small. At the peak of this massive extinction, around 13,500 BP, *global* size of human population was only around 4 million (see Table 1). The natural increase was only around 350 persons per year. This small size of global population and its small annual increase adds to the evidence that *the massive extinction of species of megafauna was most likely not caused by humans.*

People must have lived in small isolated communities with massive land areas all around them. Out of the already small number of people, women, children and older generations did not hunt. The number of hunters was, therefore, small and their best hunting equipment consisted of stone-made implements, which were hard to produce, easy to damage, hard to replace or repair, the implements efficient perhaps for hunting small pray but not for massive killing of large animals.

Efficient means of locomotion allowing for reaching new hunting grounds did not exist. Gravity assisted killing, if at all applicable, could be used only in certain geographical locations and it required a participation of a sufficiently large number of hunters. Examples of human-mediated *depletion* of a single species in some small isolated places might have happened but did it happen over larger areas and did it happen with such intensity as to cause not just the *extinction* of one or two species but the *massive extinction* of many species?

The supply of food was abundant. There was no apparent reason for killing megafauna. To cause their extinction one would need to assume a massive killing on an extraordinary large scale. Why should humans be interested in doing it? How were they were supposed to do it? Who was supposed to eat all this large amount of food?

Killing them accidently might be considered, such as killing by a regular and repeated burning of forests, but why should humans living all over the world be so determined to burn forests on a regular basis?

Australian aboriginal population is sometimes blamed for the extinction of Australian megafauna by burning forests (Miller, Fogel, Magee, Gagan, Clarke & Johnson, 2005). Aboriginal populations lived with nature and had respect for the land. Would they be so irresponsible or so little caring about their survival that they would be burning forests all around them? Would they be so determined or so careless to destroy their habitat and their food supply? Would they be able to do it?

At the time of the extinction of megafauna in Australia around 46,000 BP (Roberts, et al., 2001), the size of human population was only around 2,000 and the growth rate was approximately 0.004% (Nielsen, 2017d). The world population at that time was just over 1 million and it was at the beginning of a transition to a faster hyperbolic trajectory (see Figure 2).

With such a small and approximately stable human population in Australia, and with such large land areas all around them, would they even be able to kill so many animals as to cause their massive extinction? Sakaguchi, et al. (2013) argue that it was not fire but climate that killed Australian megafauna.

Extinction of megafauna by humans is not supported by the population data. We have to look for other possible reasons. Extinction of species is common and there is no particular reason why megafauna should be excluded from this process.

There appears to be no justification for putting so much emphasis on human impact while many other factors and forces could have contributed to the process of extinction, factors and forces much stronger than human impacts. The most obvious and much more powerful force is climate change. The extinction might have depended on the frequency of climate-related events, their intensity and their general pattern, but the resilience of species to climate change and their adaptation abilities might obscure the expected correlations. The effects of climate change depend also on geographical locations.

Some other obvious factors that might have contributed to the extinction of certain species of megafauna in certain geographical locations include the availability of refuges in certain areas, the number and the type of megafaunal species in any place, the number and the type of predators, the time-dependant access to water, the time-dependant availability of suitable vegetation, the migration of species including their interaction and competition for food resources and for shelters, the rate of natural increase (replacement efficiency) and maybe even the gestation period. We also have to consider that the extinction of megafauna was not like the total extinction of dinosaurs because many species of megafauna survived to our time.

With so many contributing factors, the problem of the extinction of certain species of megafauna might never be solved. Vast amount of data would have to be collected and analysed. However, the current huge and destructive anthropogenic impacts on the environment should not be readily extrapolated to the time when the size of human population was small, its growth was negligible and when the technology was in its primitive stages of development. *Human population dynamics does not support the postulate of the Late-Pleistocene human-assisted extinction of megafauna.*